\documentclass[aps,prl,onecolumn,groupedaddress]{revtex4}
\usepackage{graphicx}
\usepackage{epstopdf}
\usepackage{color}

\usepackage{amsmath,amsthm,amssymb,euscript,mathrsfs,epsf,epsfig}

\begin{document}

\title{Thermally triggered phononic gaps in liquids at THz scale}

\author{Dima Bolmatov$^{1}$}
\email{d.bolmatov@gmail.com, bolmatov@bnl.gov}
\author{Mikhail Zhernenkov$^{1}$}
\email{zherne@bnl.gov}
\author{Dmitry Zav'yalov$^{2}$}
\author{Stanislav Stoupin$^{3}$}
\author{Alessandro Cunsolo$^{1}$}
\author{Yong Q. Cai$^{1}$}

\affiliation{$^1$ National Synchrotron Light Source II, Brookhaven National Laboratory, Upton, NY 11973, USA}
\affiliation{$^2$ Volgograd State Technical University, Volgograd, 400005 Russia}
\affiliation{$^3$ Advanced Photon Source, Argonne National Laboratory, Argonne, Illinois 60439, USA}

\begin{abstract}
In this paper we present inelastic X-ray scattering experiments in a diamond anvil cell and molecular dynamic simulations to investigate the behavior of phononic excitations in liquid Ar. The spectra calculated using molecular dynamics were found to be in a good agreement with the experimental data. Furthermore we observe, that upon the temperature increase, a low-frequency transverse phononic gap emerges while high-frequency propagating modes become evanescent at the THz scale. The effect of strong localization of a longitudinal phononic mode in the supercritical phase is observed for the first time. The evidence for the high-frequency transverse phononic gap due to the transition from an oscillatory to a ballistic dynamic regimes of motion is presented and supported by molecular dynamics simulations. This transition takes place across the Frenkel line thermodynamic limit which demarcates compressed liquid and non-compressed fluid domains on the phase diagram and supported through calculations within the Green-Kubo phenomenological formalism. These results are crucial to advance the development of novel terahertz thermal devices, phononic lenses, mirrors, and other THz metamaterials.
\\
* d.bolmatov@gmail.com, $\dagger$ zherne@bnl.gov

\end{abstract}


\maketitle
Liquid aggregation phase results from a combination of  strong interactions and cohesive states as in solids, and large flow-enabling particle displacements as in gases \cite{ffrenkel}. Sound propagates well in liquids in the form of longitudinal phonons, however it was not evident whether transverse phonons (which propagate in solids) also exist in liquids \cite{llandau}. In a recent study a  link was established  between the positive sound dispersion (PSD) effect \cite{stanley}, also called the {\it fast or second sound}, and the origin of high-frequency transverse phononic excitations \cite{annals2015,jpcl2015}. Both these effects can be considered as a universal fingerprint of the high-frequency dynamic response of a liquid.  Each phonon mode contributes to the internal energy and heat capacity \cite{maldovan}, hence, it is important to know how many phononic states exist in a liquid of interest. Therefore, further advances in manipulation of individual phononic states in liquids and liquid-based devices is  limited by the lack of knowledge about the number of allowed states and phononic localization/delocalization effects in these systems. Despite recent advancements made in the development of phononic technologies such as topological fluid acoustics \cite{baile}, microfluidics \cite{uspal}, hypersound phononics \cite{fytas1,fytas2} and phononics in colloid-based systems at GHz scale \cite{fytas3,fytas4}, the understanding of the phononic states evolution in liquids at THz scale is still elusive.

In this work, we present the results of inelastic X-ray scattering (IXS) experiment in a diamond anvil cell (DAC) combined with molecular dynamics (MD) simulations that unveil the localization and delocalization effects in a structurally simple system, the monatomic liquid Ar. The synergy between experiment and computer simulations provided several advantages including but not limited to: {\it i)} the possibility of circumventing resolution limitations, which potentially expands the probed dynamic range toward longer distances and timescales; {\it ii)} the opportunity of discriminating the acoustic polarizations; {\it iii)} the possibility of exploring extreme thermodynamic conditions practically unattainable experimentally. Furthermore, an analytical model which predicts a low-frequency transverse phononic gaps and their evolution as a function of temperature is introduced. The model predicts both low- and high-frequency transverse phononic gaps in liquid Ar that are primarily associated with dynamic and structural changes on short and intermediate length scales ($\sim$0.2-1.4 nm).  This prediction was confirmed through calculation of correlation functions in the framework of a phenomenological linear response theory, namely the Green-Kubo formalism \cite{green,kubo}. The experimental data are consistent with MD simulations and confirm the theoretical predictions. The appearance of low-frequency transverse phononic gaps is a result of symmetry breaking due to the interaction of phonons that can be created or annihilated making materials {\it anharmonic} over the heat production. Our results show that high-frequency phononic gaps (both transverse and longitudinal) appear on picosecond time scale at elevated temperatures. This point is of particular relevance, since the manipulation of phononic gaps in the so called phononic crystals \cite{maldovan} is usually achieved through structural engineering. This is of crucial importance to implement heat management, since heat transfer in insulators uses phonons as carriers. In this work, it is shown that the phononic gap manipulation can be achieved even in simplest materials through a suitable tuning of thermodynamic conditions. This reflects into a tuning of the atomic dynamics and structure. These can be monitored either through the evolution of the velocity-velocity autocorrelation function or the pair distribution function. Tuning of the atomic dynamics and thermally triggered localization of low- and high-frequency propagating phononic modes  are the key factors for the advancement of technologies based on the sound control and manipulation at the THz regime. We anticipate that our findings will  advance terahertz thermal devices and THz techniques.
\begin{figure*}[htp]
  \centering
 \begin{tabular}{cc}
     \includegraphics[width=170mm]{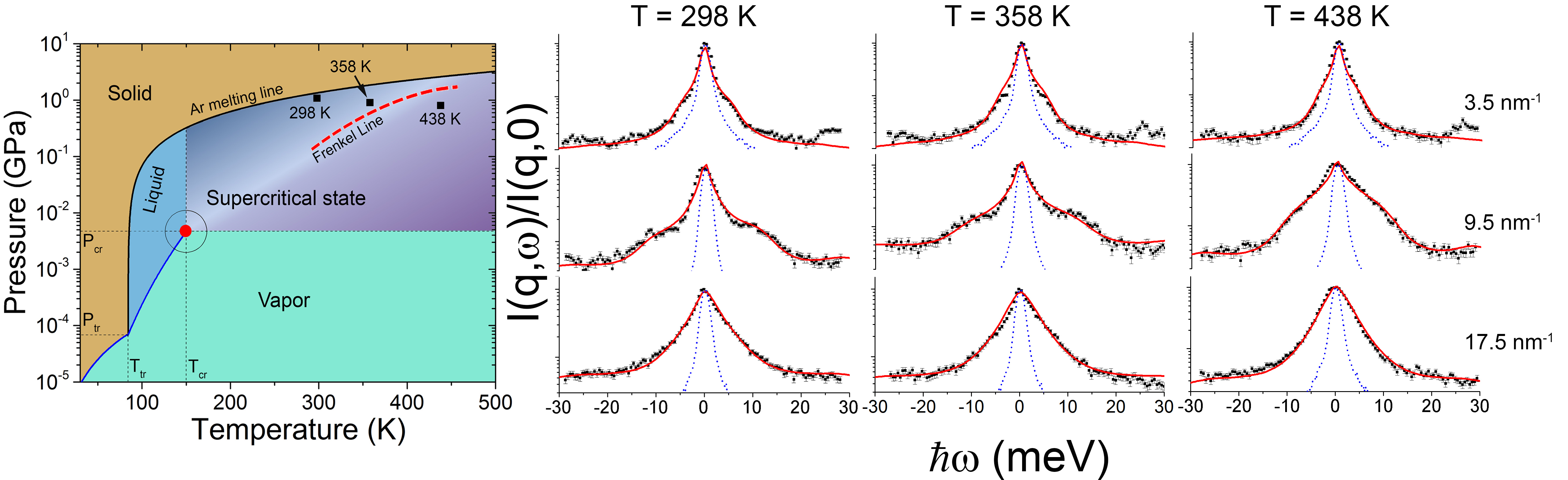}
  \end{tabular}
\caption{ {\bf Ar pressure-temperature phase diagram and IXS spectra at different thermodynamic conditions}. Inelastic X-ray scattering spectra of Ar at T = 298 K, P = 1.08 GPa in the rigid liquid phase (Left panel), at T = 358 K, P = 0.9 GPa in the "mixed" phase at the Frenkel line (Central panel) and at T = 438 K, P = 0.8 GPa below the Frenkel line in the non-rigid fluid phase (Right panel). The thermodynamic points of three reported measurements are indicated on the phase diagram. The experimental data (black squares) with error bars signifying one standard deviation are reported together with the MD simulation (red solid curves). The MD curves are presented after convolution with the IXS spectrometer resolution function, which has almost a Lorentzian shape (FWHM $\sim$2 meV) and is denoted by the dashed blue line. The resolution profile was measured experimentally for each IXS analyzer. The $q$-resolution of the spectrometer was $\simeq$2 nm$^{-1}$ and the MD simulated spectra had the corresponding $q$-values matching the $q$ of each IXS analyzer. The corresponding $q$-values are shown to the right.}
\label{spectra}
\end{figure*}
\section{Results}
Below, we introduce the effective Hamiltonian \cite{bolsruni} where the problem of strong interactions is tackled from the outset
\begin{equation}
\label{EHam}
\begin{aligned}
H[\varphi_q] =\frac{1}{2}\sum_{0\leq\omega_q^{l,t,t}\leq\omega_{\rm D}}[\pi_q^{l}\pi_{-q}^{l}+\pi_q^{t}\pi_{-q}^{t}+\pi_q^{t}\pi_{-q}^{t}]+\\
\sum_{0\leq\omega_q^l\leq\omega_{\rm D}}\left[\frac{\omega_q^2}{2}\varphi^l_q\varphi^l_{-q}\right]+ \sum_{\omega_{\rm F}\leq\omega_q^{t,t}\leq\omega_{\rm D}}\left[\frac{\omega_q^2}{2}(\varphi^t_q\varphi^t_{-q}+\varphi^t_q\varphi^t_{-q})\right]
\end{aligned}
\end{equation}
where $q$ is a multiindex $\{q_l,q_t,q_t\}$ and the collective canonical coordinates $\pi^\alpha_q$ and $\varphi_q^\alpha$ ($\alpha=l,t,t$) are defined by: $\varphi_q^\alpha= \sqrt{m} \sum_{j=1}^{N} e^{{\texttt i}L (j\cdot q) }x^\alpha_j$
and $\pi^\alpha_q = \dot{\varphi}_q^\alpha$. Here, $x_j^\alpha$  is space coordinate of an atom in a lattice sitting in a vertex labelled by the multiindex. $l$ and $t$ stand for the longitudinal and transverse phononic polarizations, respectively. $\omega_{\rm D}$ is the Debye frequency. $\omega_{\rm F}$  puts a lower bound on the oscillation frequency of the atoms and can be derived from the viscosity $\eta$ and shear modulus $G_{\infty}$ of a liquid \cite{bolsrph,boljap,bolprb}
\begin{equation}
\omega_{\rm F}(T)=\frac{2\pi}{\tau(T)}=\frac{2\pi G_{\infty}}{\eta(T)}
\label{maxwell}
\end{equation}
where $\tau=\frac{\eta}{G_{\infty}}$ is the Maxwell's relation and $\tau$ is the relaxation time. This Hamiltonian predicts
the low-frequency transverse phononic gaps in a liquid spectrum ($\omega_{\rm F}\leq\omega_q^{t,t}\leq\omega_{\rm D}$, see the last term in Eq. (\ref{EHam}))
which is a result of a symmetry breaking due to interaction of phonons \cite{annals2015,bolsruni}. Viscosity $\eta(T)$ drops down upon temperature rise resulting in the transverse phononic
gap growth  due to the Frenkel frequency $\omega_{\rm F}$ (see Eq. (\ref{maxwell})) increase. An increase in temperature leads to the progressive disappearance of the high-frequency transverse phononic modes. This leads to $\omega_{\rm F}\xrightarrow{T}\omega_{\rm D}$ and, hence,  $c_V=\left(\frac{1}{N}\frac{\partial E}{\partial T}\right)_{\rm V}$ ($H\varphi=E\varphi$): 3$k_{\rm B}\xrightarrow{T}$2$k_{\rm B}$. $c_{V}=2k_{\rm B}$ (when $\omega_{\rm F}=\omega_{\rm D}$)  is the new thermodynamic limit \cite{annals2015,jpcl2015,bolsrph,bolnature}, dubbed the Frenkel line thermodynamic limit in the previous works \cite{annals2015,jpcl2015}. Further, we will present experimental, MD simulations results and calculation of correlation functions in the framework of the Green-Kubo  phenomenological formalism  as a test of our theoretical predictions.
\begin{figure}[htp]
  \centering
 \begin{tabular}{cc}
     \includegraphics[width=75mm]{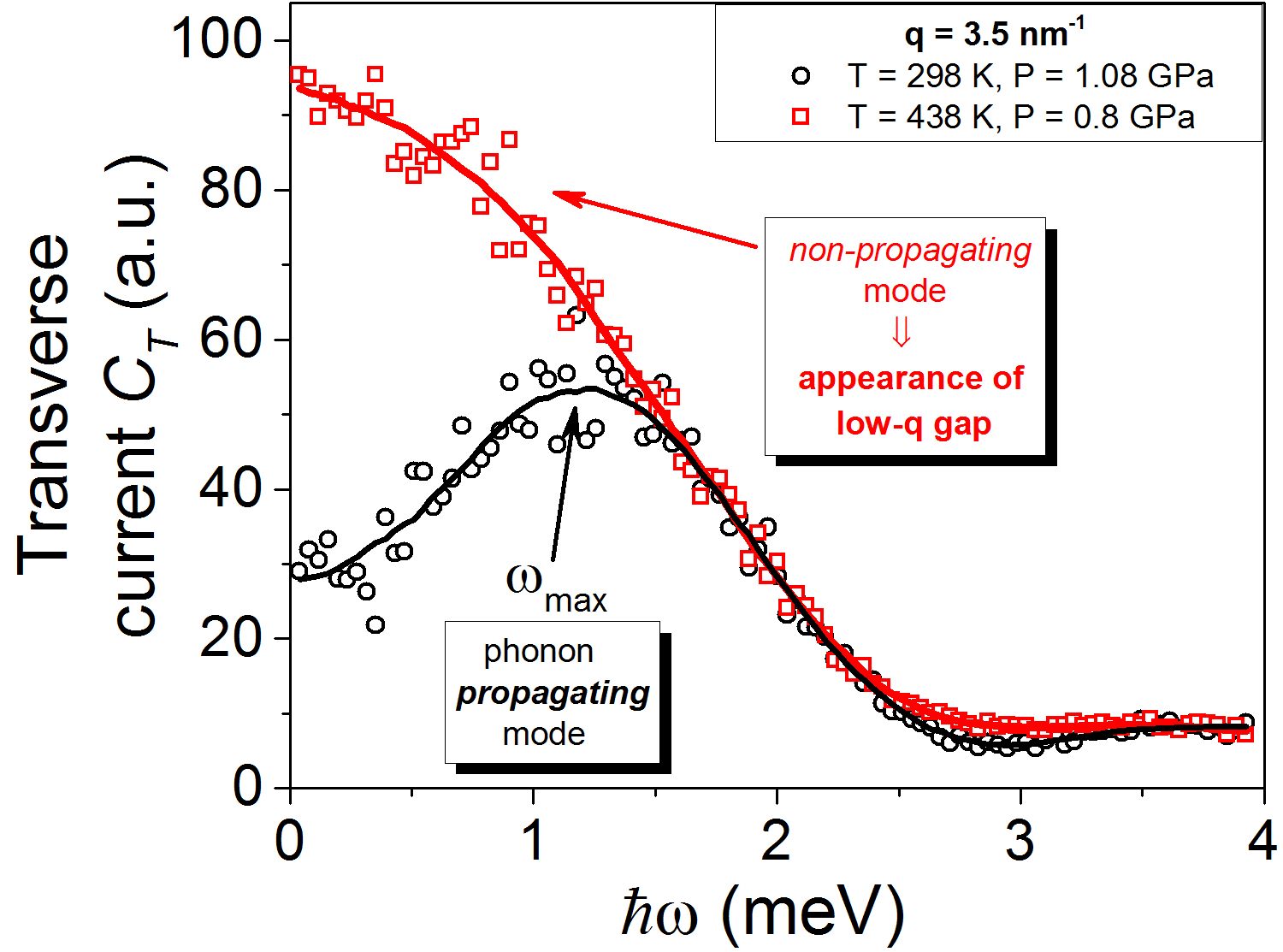}
  \end{tabular}
\caption{ {\bf Transverse current autocorrelation function $C_t$ above and below the Frenkel line calculated from MD simulations}. A clear peak (see the black line) indicates existence of transverse phononic propagating mode at T=298 K. The absence of the peak (see the red line, T=438 K) provides a compelling evidence for the emergence of the low-$q$ thermally-triggered phononic gap. Solid lines are guide to eyes only.}
\label{current}
\end{figure}

Selected spectra  normalized to their respective intensity for liquid Ar, measured by inelastic X-ray scattering (IXS)  in a diamond anvil cell (DAC), are depicted in Fig. (\ref{spectra}) at three thermodynamic points: at T = 298 K, P = 1.08 GPa in the rigid liquid phase (Left panel), at T = 358 K, P = 0.9 GPa in the "mixed" phase at the Frenkel line (Central panel) and at T = 438 K, P = 0.8 GPa below the Frenkel line in the non-rigid fluid phase (Right panel). The location of the experimental points with respect to the Frenkel line on the P-T phase diagram (see Fig. \ref{spectra}) is consistent with the position of the Frenkel line discovered in previous X-ray diffraction measurements \cite{fline2015}. The high quality data clearly proves the existence of longitudinal phononic modes, which appear as peaks or shoulders in the lower $q$-range. The reason why peaks associated to transverse modes are not visible in the measured lineshapes is that the spectrum $S(q,\omega)$ couples primarily with longitudinal modes only. Transverse modes become visible exclusively when longitudinal and transverse polarizations are coupled, which is not the case of a noble gas. 

The calculated dynamic structure factor from MD simulations  with the sum of a delta function for the elastic component and convoluted with the experimental resolution function is in a good agreement with the experimental $S(q,\omega)$ (see Fig. (\ref{spectra})). MD simulations' code used to model the IXS spectra (except for the elastic line, the experimental background and the balance factor, see {\bf Methods} section for details) is free of adjustable parameters. This is a fundamentally different approach from fitting models such as the customarily used damped harmonic oscillator (DHO) model.  Both simulated and measured spectra bear evidence of well defined inelastic shoulders whose position seems clearly increasing with $q$ at low/moderate exchanged momenta. It can also be noticed that the lowest $q$ IXS spectra exhibit additional high frequency peaks in the upper extreme of the probed frequency range (see Fig. (\ref{spectra})), these peaks, not observable in the simulated spectra are to be ascribed to the phonons excitations dominating the scattering from the diamond DAC windows. It has to be noticed that the determination of the dispersion curves, i.e. of the $q$-dependence of inelastic shift of the measured spectra is not straightforward. This is mainly due to both instrumental resolution limitation and the highly damped nature of the collective modes in the spectrum. Consequently, a detailed modeling of the measured lineshape accounting for the contribution of the experimental resolution would be required. This modeling has been thoroughly discussed in Ref. \cite{jpcl2015}.

\begin{figure}[htp]
  \centering
 \begin{tabular}{cc}
     \includegraphics[width=75mm]{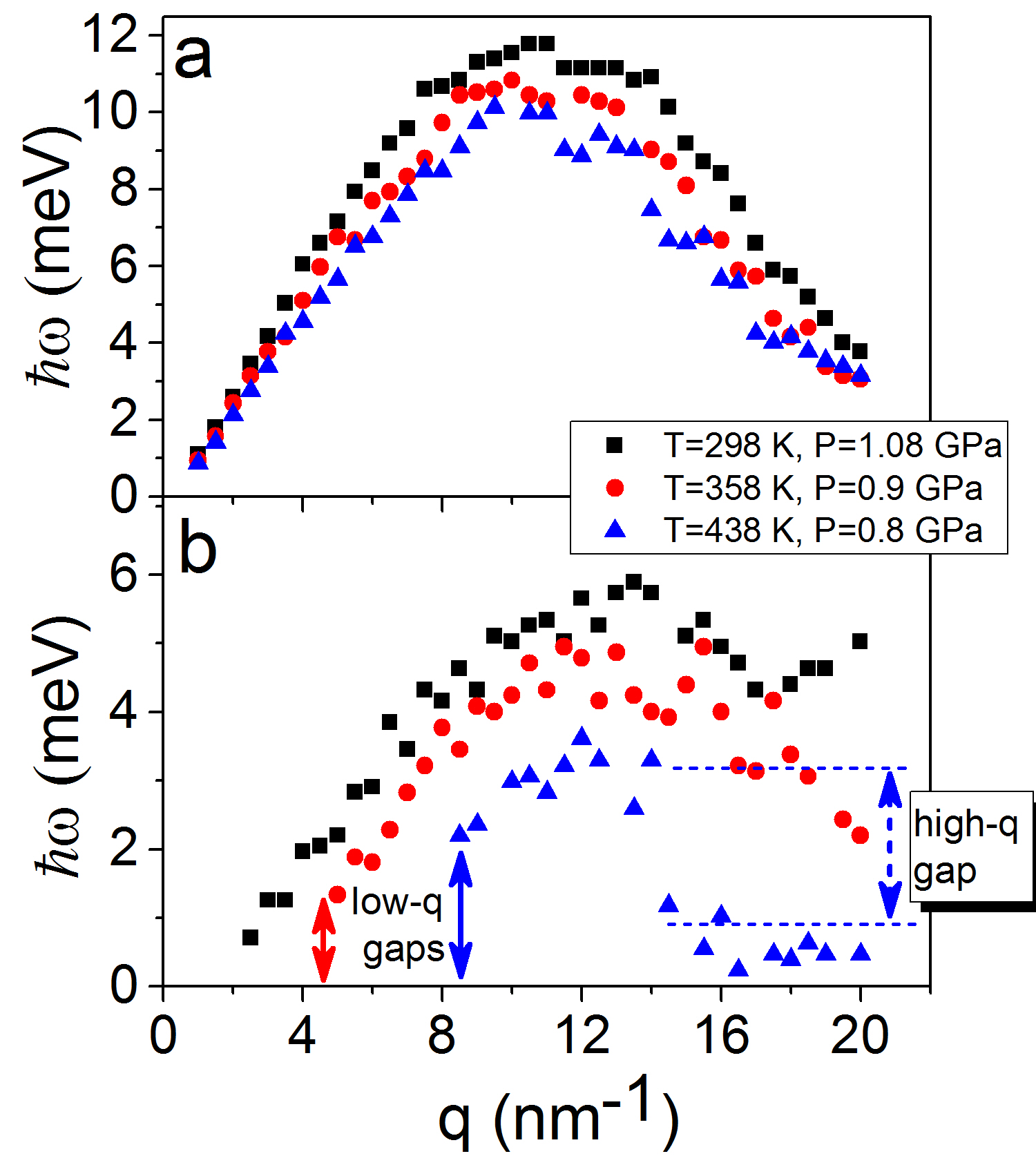}
  \end{tabular}
\caption{ {\bf Evolution of  longitudinal and transverse phononic excitations calculated from MD simulations at experimental conditions}. {\bf (a,b)} Longitudinal and transverse $\hbar\omega$ dispersions are derived from the maxima of the corresponding current autocorrelation functions. {\bf (b)} Emergence of low-frequency transverse  phononic gap at T=358 K and high-frequency phononic gap at T=438 K are indicated.}
\label{dis}
\end{figure}

Density fluctuations in compressed liquids reduce the intensity of the Bragg diffraction peaks and, as a result, increase the diffuse scattering between these peaks \cite{monaco1}.  Liquids
sustain medium range order correlations \cite{bolstr1,bolstr2} but the translationally invariant symmetry is broken \cite{bolsruni}. Therefore, liquids do not have periodicity, and the first and second Brillouin zones are not well defined as a result. The broken translational symmetry induces a boundary smearing between Brillouin zones. Hence, the transverse phononic modes are heavily damped at higher $q$ values (see Fig. (\ref{spectra})) but can clearly be evidenced from the corresponding MD simulations results. 

\begin{figure}[htp]
  \centering
 \begin{tabular}{cc}
     \includegraphics[width=75mm]{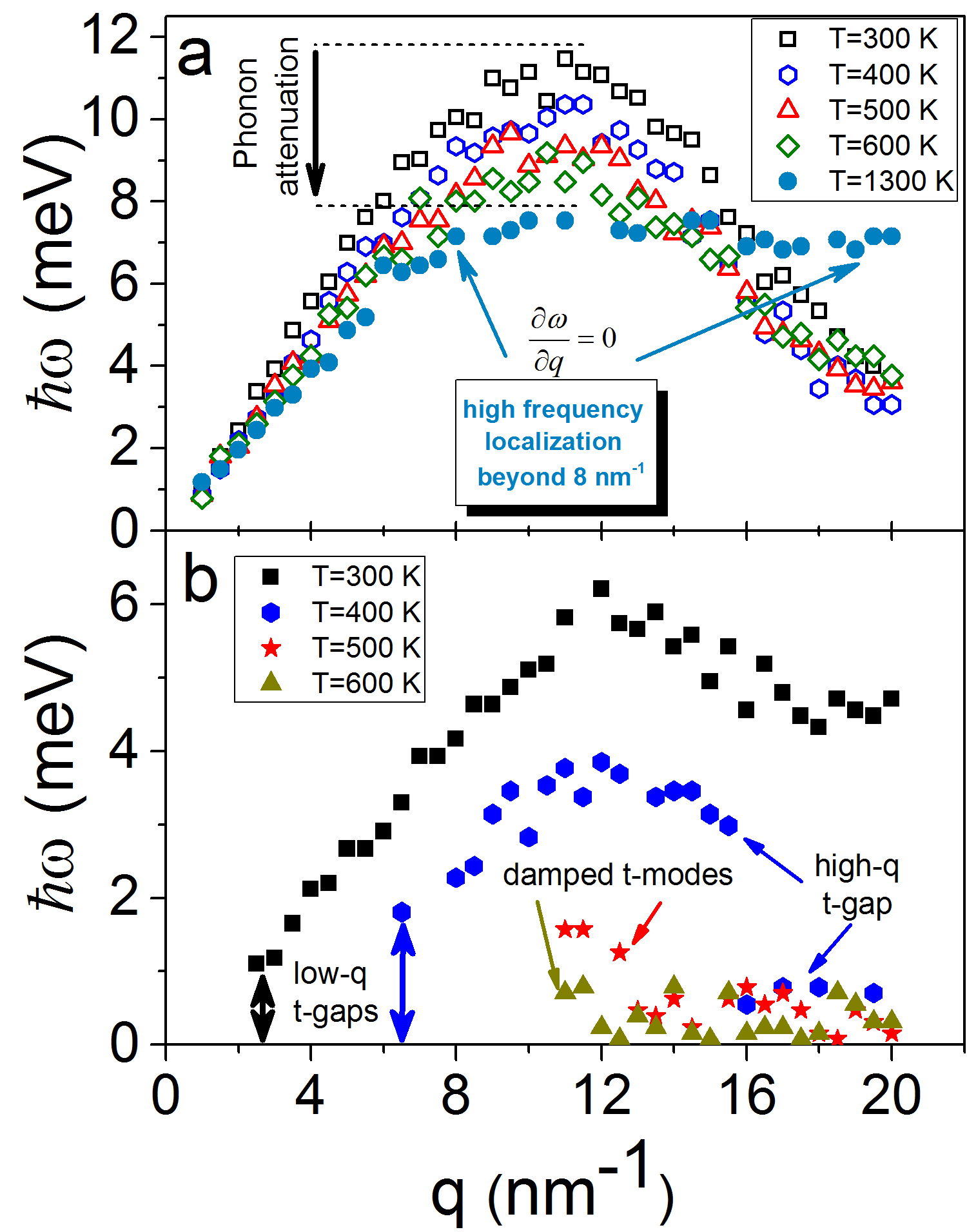}
  \end{tabular}
\caption{ {\bf Temperature variations of longitudinal and transverse phononic modes calculated from MD simulations at fixed pressure}. {\bf (a,b)}  Longitudinal and transverse $\hbar\omega$ dispersions are derived from the maxima of the corresponding current autocorrelation functions. {\bf (a)} Longitudinal phononic modes experience attenuation. {\bf (b)} Emergence of low-frequency transverse phononic gap at low temperatures and high-frequency transverse phononic gap at elevated temperatures. Transverse phononic excitations become heavily overdamped upon crossing the Frenkel line ($\sim$ 358 K, P=0.9 GPa) and ($\sim$ 400 K, P=0.8 GPa). Below the Frenkel line transverse phononic modes do not propagate any more {\bf (b)} and high-$q$ longitudinal modes become strongly localized $\left(\frac{\partial\omega}{\partial q}=0\right )$ as temperature rises {\bf (a)}.}
\label{dis1}
\end{figure}

In order to unveil regimes of phonon propagation and localization effects in liquids we calculate longitudinal and transverse autocorrelation functions. The transverse autocorrelation function $C_{t}$ above and below the Frenkel line is depicted in Fig. (\ref{current}) providing a compelling evidence for the emergence of the thermally triggered transverse phononic  gap. To analyze this phenomenon in detail, we present the dispersion relations at experimental conditions (see Fig. (\ref{dis})) and beyond (see Fig. (\ref{dis1})). Longitudinal and transverse $\hbar\omega$ dispersions are derived from the maxima of the corresponding current autocorrelation functions. Figs. (\ref{dis}.b-\ref{dis1}.b) display the evolution of the low-frequency transverse phononic gaps. The emergence of the low-$q$ transverse phononic gap occurs due to symmetry breaking in phonon interactions \cite{bolsruni}  and reflected in the last term of the effective Hamiltonian (\ref{EHam}). The transverse phononic gap grows over the heat production and becomes heavily overdamped upon crossing the Frenkel line (see Fig.{\ref{dis1}.b}).  The inability to support low-frequency transverse phononic excitations (low-energy cutoff) is a manifestation of the absence of the long-range  pair correlations in liquids and {\it vise versa}. An
increase in temperature leads to the disappearance of both the high-frequency transverse phononic modes (see Fig. \ref{dis}) and progressively the medium-range pair correlations, which we show in detail in the next section.
\begin{figure*}[htp]
  \centering
 \begin{tabular}{cc}
     \includegraphics[width=150mm]{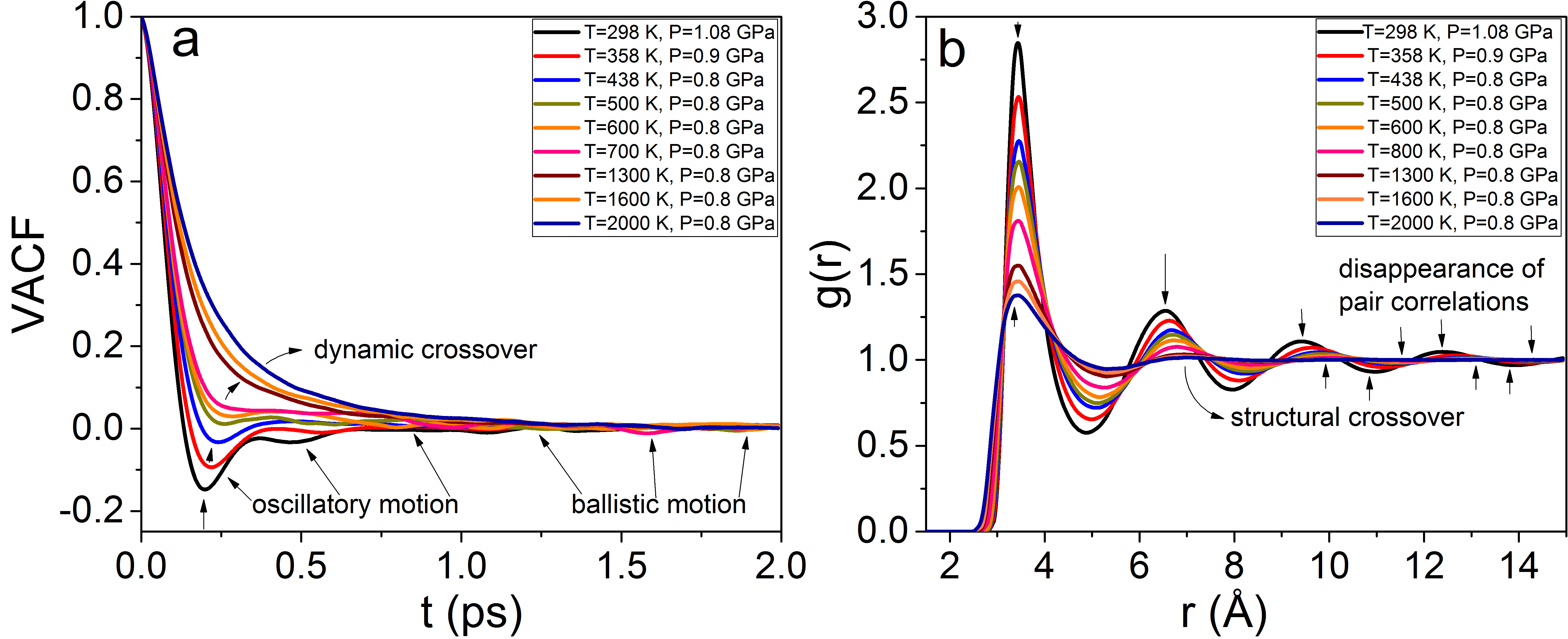}
  \end{tabular}
\caption{ {\bf Evolution of dynamic (VACF) and structural ($g(r)$) correlation functions}. {\bf (a)} VACF and $g(r)$ {\bf (b)} pair correlation functions evolve consistently at the same thermodynamic conditions: both at experimental conditions (T=298 K, T=358 K and T=438 K) and beyond. VACF undergoes the dynamic crossover {\bf (a)} while $g(r)$ undergoes the structural crossover {\bf (b)} showing the interconnection between dynamics and structure in the liquid Ar in the framework of the Green-Kubo formalism.}
\label{corr}
\label{correlations}
\end{figure*}
Along with the existence of the low-q transverse phononic gap we observe the emergence of the high-frequency transverse gap in two different thermodynamic points: (T=438 K, P=0.8 GPa; see Fig. (\ref{dis}.b))) and (T=400 K and T=500 K, P=0.8 GPa; see Fig. (\ref{dis1}.b)). The high-frequency phononic gap has recently been  observed in the vicinity of the freezing point \cite{sanz}. The emergence of high-q transverse phononic gaps in the liquid and supercritical states is an unexpected and surprising result in view of structural simplicity of monatomic liquid Ar. We ascribe the appearance of the high-$q$ gaps  to the change in the atomic dynamics and pair correlations upon crossing the Frenkel line. Below the Frenkel line the transverse phononic modes are heavily overdamped and, therefore, are not supported in the non-rigid fluid regime. As temperature increases in the non-rigid fluid regime (see Fig. (\ref{dis1}.a)) the longitudinal phononic modes attenuation takes place with strong phononic localization $\left(\frac{\partial \omega}{\partial q}=0\right)$ , which has been unanticipated. As a complementary test for our theoretical predictions, we calculate below correlation functions in the framework of the phenomenological Green-Kubo formalism at experimental thermodynamic conditions and beyond.

Green and Kubo demonstrated \cite{green} that {\it phenomenological} transport coefficients could be introduced as integrals over a time-correlation function in thermal equilibrium. The relaxation time and dissipation of fluctuations have the same origin toward the equilibrium when a system is out of equilibrium due to an external force. Therefore,  their scale are determined by the same transport coefficients. Here, we calculate the velocity-velocity autocorrelation function (VACF) at experimental conditions and beyond (see Fig. (\ref{corr}.a)). If the velocity vector for a system of atoms is $\mathbf{v}(t)$, then the VACF $Z(t)$ can  be expressed as
\begin{equation}
\displaystyle Z(t) =\frac{\left\langle\mathbf{v}(0)\cdot\mathbf{v}(t)\right\rangle}{\left\langle\mathbf{v}(0)\cdot\mathbf{v}(0)\right\rangle}
\end{equation}
with $D\propto Z(t)$, where $D$ is the diffusion constant. The rapid decrease of VACF implies the decrease in the pair correlations of atomic motion along the trajectories of the atoms. In a compressed liquid the VACF is not a simple exponential but has a negative part as its characteristic solid-like feature up to 1--1.5 ps.  $Z(t)$ becomes shallow at longer times exhibiting a gas-like behavior. When the atomic motions become  oscillatory, the VACF describes the oscillations because the velocity of the atoms self-correlates in a solid-like manner. Similarly, if the velocity of atoms has a definite direction, then the VACF gradually raises in its magnitude revealing the diffusive behavior of the atoms. In Fig. (\ref{corr}.a) we depict the evolution of VACF with temperature variations and evidence the dynamic crossover \cite{gallo}. Secondly, we calculate pair distribution function $g(r)$ ($E\propto g(r)$, where $E$ is the internal energy, $H\phi=E\phi$) at the same thermodynamic conditions as for VACF. This function describes the distribution of distances between atomic pairs subjected to a given volume \cite{bolstr1,bolstr2}. In Fig. (\ref{corr}.b),  we show that liquid Ar undergoes the structural crossover which corresponds to the qualitative change of atomic structure, the continuous transition of the substance from the {\it rigid liquid} structure (pair correlations presented on an intermediate length scale) to the {\it non-rigid fluid} structure (pair correlations preserved on a short length scale only). 

We now discuss the origin of the structural crossover (see Fig. \ref{corr}.b), and relate it to the change of dynamics (see Fig. \ref{corr}.a) of the liquid state. $g(r)$ peaks decrease rapidly due to the exponential decrease of the relaxation time $\tau$ (also see \cite{bolstr1}). In the {\it non-rigid fluid} regime where the oscillatory component of motion (see Fig. \ref{corr}.a) and the medium-range pair correlations are no longer present (see Fig. \ref{corr}.b), pair correlations are less sensitive to temperature increase because the dynamics is already randomized as in a gas phase. This implies that the system reached the Frenkel line thermodynamic limit \cite{jpcl2015}: $\omega_{\rm F}\xrightarrow{T}\omega_{\rm D}$ (see Eq. (\ref{EHam})) giving $c_V=\left( \frac{1}{N}\frac{\partial E}{\partial T}\right)_{\rm V}$=2$k_{\rm B}$ (see Eq. (\ref{EHam})). This picture supports the relationship between the dynamics (phononic excitations), structure (pair correlations) and thermodynamics (heat capacity), which is consistent with the temperature evolution on the P-T phase diagram.

Importantly, the Frenkel line is a smooth and continuous
thermodynamic boundary. This boundary begins in the neighborhood of the critical point and extends into the supercritical area up to very high P-T conditions \cite{annals2015, jpcl2015, fline2015}. Strictly speaking, however, at the Frenkel line $c_V\simeq 2k_{\rm B}$; not exactly $c_V=2k_{\rm B}$ due to anharmonic effects ( $\alpha\neq 0$) \cite{bolprb}, where  $\alpha$ is the thermal expansion coefficient. Thus, $c_V=2k_{\rm B}$ is the new thermodynamic limit \cite{annals2015, jpcl2015, fline2015},  in contrast to what used to be previously believed,  but the thermodynamic boundary namely the Frenkel line which is smeared out in its vicinity, where major thermodynamic quantities undergo their maxima or minima \cite{gallo}.
\section{Discussion}
The interatomic interactions in a liquid are strong, yet the structure is disordered which substantially hampers calculations of the phononic properties. It is noteworthy, perturbation theories and their analytic expressions which bear "small parameters" such as virial expansion, Percus-Yevick approach, Meyer expansion and others  may describe systems with weak interactions only. Thus, perturbation theories  fail to describe systems with strong interactions like compressed/rigid liquids. As a result, both dynamic (see Fig. (\ref{corr}.a)) and structural ((see Fig. (\ref{corr}.b)) crossovers, which we revealed and interconnected in this work, have not been observed before. Here, the presented experimental and MD simulations results are explained  within the framework of unified phonon-based approach \cite{annals2015} where the problem of strong interactions is tackled from the outset (see Eq. (\ref{EHam})). 

In this work, we unveiled regimes of phonon propagation and localization effects mapping the phononic excitations landscapes across the Frenkel line. The statistical mechanics' analysis of the presented data is based on temperature variations. Noteworthy, the  {\it intra} (not {\it inter}) pair distribution function $g(r)$ exhibits linear dependence on pressure variation making it impossible to detect any thermodynamic boundaries in the supercritical phase \cite{santoro}. Here, in the framework of the effective Hamiltonian (see Eq. (\ref{EHam})) we predicted the existence of the low-frequency transverse phononic gaps that progressively grow as temperature rises (see the last term in Eq. (\ref{EHam}). The validity of this prediction was reaffirmed through calculation of the dispersion relations shown in Figs. (\ref{dis})-(\ref{dis1})). The appearance of the high-frequency transverse phononic gaps in structurally simple system such as liquid Ar is an unexpected result. The origin of this result was revealed by studying the evolution of correlation functions within the Green-Kubo formalism. The velocity-velocity autocorrelation function has an oscillatory component of motion (picosecond scale) while the $g(r)$ possesses medium-range correlations (0.2-1.4 nm) above the Frenkel line.  Upon crossing it, the oscillatory component vanishes and the motions become purely ballistic (see Fig. (\ref{corr}.a))  and, at the same time, the medium-range pair correlations  disappear  (\ref{corr}.b)) triggering the high-frequency phononic gaps at the THz scale.

The unveiled regimes of phononic propagation and localization effects in liquid Ar and the existence of both low- and high-frequency phononic gaps at THz scale support the forthcoming developments of sound control and heat management through phononic gaps manipulations at meso- and nanoscales. In particular, the phononic gap manipulation can be realized by immersing nanoparticles  (with  directional bonds between them) in liquids and aqueous solutions of interest \cite{gang}. Tuning size, shape and material of the nano-blocks is the key factor for the emerging technology of phononic harvesting nano-clusters and for the phononic gaps engineering with controllable sound output. Therefore, we expect that the reported results will advance the field of matematerials at the THz scale.

\section{Methods}
{\bf IXS experiments.} We performed high pressure/high temperature inelastic x-ray scattering measurements using a BX90 Diamond Anvil Cell \cite{kantor} at the IXS beam line 30-ID of the Advanced Photon Source (APS), Argonne National Laboratory. The DAC was used in combination with tungsten-carbide seats and full diamond anvils with a 500 $\mu$m culet size.  250 $\mu$m-thick rhenium gasket was  pre-indented to a thickness of about 90 $\mu$m. A hole with a diameter of about 220 $\mu$m was drilled in the middle of the pre-indented area. Conventional resistive heating was used to heat the sample. The $^{\rm 40}$Ar was loaded using a COMPRES/GSECARS gas-loading system at APS \cite{compress} up to initial pressure of 1.08 GPa. A ruby sphere was used for the pressure calibration \cite{sandeep}. The sample was measured at the photon energy of 23.724 keV; Following every temperature change, the DAC was allowed to equilibrate for, at least, 15 minutes before the IXS spectrum was collected.

The MD simulation is intrinsically unable to probe the extremely long time response of the fluid ($\omega\cong 0$) mainly concentrating around the elastic position.  In order to perform a reliable comparison between measured and simulated spectra, an elastic $\propto \delta(\omega)-$function was added. The elastic term accounts for all parasitic intensity effects contributing to the measured elastic intensity, {\it e.g.} empty can contribution.
The MD lineshapes were corrected by a detailed balance factor which governs the positive and negative phononic energy intensities \cite{sette1}. The MD lineshapes combined with the elastic term were convoluted with the experimentally measured instrumental resolution functions. Furthermore, the experimental IXS background contribution was taken into account when comparing the MD simulations and the IXS data. The elastic intensity and background level were treated as variables whose optimized values provided the best fit to the experimental spectra.  The resulting MD lineshapes are reported as red solid lines in Figure (\ref{spectra}).

{\bf MD simulations.} To calculate dispersion relations and correlation functions, we have used LAMMPS simulation code to run a Lennard-Jones (LJ, $\varepsilon/k_{\rm B}$=119.8 K, $\sigma$=3.405) fluid fitted to Ar properties with 32678 atoms in the isothermal-isobaric (NPT) ensemble.

\section{Acknowledgements} 
The work at the National Synchrotron Light Source-II, Brookhaven National Laboratory, was supported by the U. S. Department of Energy, Office of Science, Office of Basic Energy Sciences, under Contract No. DE-SC0012704. Synchrotron experiment was performed at 30-ID beamline, Advanced Photon Source (APS), Argonne National Laboratory. Use of the Advanced Photon Source was supported by the U. S. Department of Energy, Office of Science, Office of Basic Energy Sciences, under Contract No. DE-AC02-06CH11357. We thank Bogdan M. Leu and Ayman H. Said for their support during the experiment at Sector 30 at APS, and Sergey N. Tkachev for his help with the GSECARS gas loading system. We thank staff of GSECARS at Advanced Photon Source for providing the diamond anvil cell and resistive heating system. We are grateful to Oleg Gang and Yugang Zhang for stimulating discussions. 

\section{Author contributions}
D.B., M.Z., A.C. and Y.Q.C. designed the research. M.Z., D.B., S.S., A.C. and Y. Q. C. prepared the high pressure cell and performed resistive heating experiment. D.B., D.Z., A.C., Y.Q.C. and M. Z. performed the numerical simulations. D.B. and D.Z. performed MD simulations.  D.B. and M.Z. wrote the manuscript. All authors discussed the results and commented on the manuscript.

\section{Competing financial interests}

The authors declare no competing financial interests.

\end{document}